\documentclass[%
 aip,
 amsmath,amssymb,
 reprint,%
]{revtex4-1}

\usepackage{graphicx}
\usepackage{dcolumn}
\usepackage{bm}

\usepackage[utf8]{inputenc}
\usepackage[T1]{fontenc}
\usepackage{mathptmx}
\usepackage{etoolbox}
\usepackage{xcolor}

\newcommand{\alps}{\texttt{ALPS}}
\newcommand{\nhds}{\texttt{NHDS}}
\newcommand{\plume}{\texttt{PLUME}}

\newcommand{\Alfven}{Alfv\'{e}n }

\newcommand{\V}[1]{\boldsymbol{#1}}
\renewcommand{\vec}[1]{\boldsymbol{#1}}
 
\newcommand{\xhat}{\mbox{$\hat{\mathbf{x}}$}}

\newcommand{\zhat}{\mbox{$\hat{\mathbf{z}}$}}

\newcommand{\change}[1]{{#1}}

\makeatletter
\def\@email#1#2{%
 \endgroup
 \patchcmd{\titleblock@produce}
  {\frontmatter@RRAPformat}
  {\frontmatter@RRAPformat{\produce@RRAP{*#1\href{mailto:#2}{#2}}}\frontmatter@RRAPformat}
  {}{}
}%
\makeatother
\begin{document}

\preprint{AIP/123-QED}

\title[Plasma Dielectric Response for Arbitrary VDFs]
{The Dielectric Response of Plasmas with Arbitrary Gyrotropic Velocity Distributions}
\author{K.G. Klein}
 \email{kgklein@arizona.edu}
\affiliation{
Lunar and Planetary Laboratory, University of Arizona, Tucson, AZ 85721, USA
}
\author{D. Verscharen}%
\affiliation{ 
Mullard Space Science Laboratory, University College London, Dorking RH5 6NT, UK
}%

\date{\today}

\begin{abstract}
Hot and tenuous plasmas are frequently far from local thermodynamic equilibrium, necessitating sophisticated methods for determining the associated plasma dielectric tensor and normal mode response.
The Arbitrary Linear Plasma Solver (\alps) is a numerical tool for calculating such responses of plasmas with  arbitrary gyrotropic background velocity distribution functions (VDFs). 
In order to model weakly and moderately damped plasma waves accurately, we have updated the code to use an improved analytic continuation enabled by a polynomial basis representation.
We demonstrate the continuity of solutions to the linear Vlasov--Maxwell dispersion relation between bi-Maxwellian and arbitrary VDF representations and evaluate the influence of VDF structure on mode polarization and wave power emission and absorption.
\end{abstract}

\maketitle

\section{Introduction}
\label{sec:intro}

Numerical solvers to the linear Vlasov--Maxwell dispersion relation have been extensively used to study the characteristics of linear plasma normal modes, in particular their growth or damping rates depending on  plasma conditions \cite{Gary:1993,Verscharen:2013a,Klein:2018,Klein:2021,McManus:2024}.
However, typical linear plasma dispersion solvers assume a particular analytical form for the underlying background velocity distribution (VDF) $f_{0,s}(\V{v})$, e.g.\ a bi-Maxwellian (\texttt{WHAMP} \citep{Roennmark:1982}, \texttt{PLUME} \citep{Klein:2025-PLUME}, \texttt{NHDS} \citep{Verscharen:2018a}, or \texttt{BO} \citep{Xie:2019}) or a bi-$\kappa$ (\texttt{DSHARK} \citep{Astfalk:2015}) distribution.
Such an assumption enables the determination of the linear dispersion relation through the evaluation of a closed equation in terms of known special functions that simplify the required integrals over velocity space.
However, many space and astrophysical plasmas, and collisionless plasmas generally, are not well represented by bi-Maxwellian, or even bi-$\kappa$, distributions.
The plasma systems can be in a state far from local thermodynamic equilibrium, c.f. reviews \change{by \citet{Marsch:2012}} and \citet{Verscharen:2019}.
These departures require a more sophisticated treatment of the dispersion relation.

The Arbitrary Linear Plasma Solver (\alps), is an MPI parallelized numerical code written in \texttt{FORTRAN90} that solves the Vlasov--Maxwell dispersion relation for a hot, magnetized plasma\cite{ALPS:2023}. 
\alps\ allows for any number of particle species with arbitrary gyrotropic background distribution functions supporting normal modes with any direction of propagation with respect to the background magnetic field, and can include the effects of special relativity in the dielectric plasma response.

Instead of using parameterized values for a collection of analytical functions, \alps\ uses as input the phase-space density on a discrete grid of parallel and perpendicular momentum $f_{0,s}(p_\perp,p_\parallel)$, which in the non-relativistic case is directly equivalent to a velocity grid $f_{0,s}(v_\perp,v_\parallel)$).
\alps\ evaluates the dispersion relation through a direct numerical integration of the gradients and other functions of of $f_{0,s}$; see Eqns. 2.9 and 2.10 of \citet{Verscharen:2018}.
This method has been applied to spacecraft data from both MMS \citep{Jiang:2022,Jiang:2024,Afshari:2024} and Wind \citep{Walters:2023}, showing significant deviations of wave behavior from predictions calculated using simple analytical functions.
It has been applied to the study of a variety of kinetic numerical simulations with significant departures for Maxwellian VDFs  \cite{Zhang:2025,Fitzmaurice:2025,Schroder:2025}.

The initial \alps\ implementation is described by \citet{Verscharen:2018}.
\change{Recent additions to the code}  include the calculation of the associated eigenfunctions and a partitioning of the species heating rates into the component mechanisms, e.g. Landau, Transit Time, and Cyclotron damping, \change{as well as the option to treat selected plasma components as bi-Maxwellians for numerical expediency. 
These changes are discussed in \S~\ref{sec:code}.}
\change{The original code struggled with the evaluation of the Landau contour integral essential for characterizing damped solutions for some VDFs with strong non-Maxwellian features.
We present updates that enable a more accurate representation in this evaluation for moderately damped solutions for such VDFs.}
We explore the impact of the updated representation on weakly and moderately damped modes in \S~\ref{sec:method}.
We demonstrate the continuity of solutions between bi-Maxwellian fits and the \alps\ direct calculation, as well as the changes for the underlying eigenfunctions and damping rates, in \S~\ref{sec:compare}.
Details of the underlying numerics are found in Appendices \ref{app:NHDS}, \ref{app:eigen}, and \ref{app:cheb}.

\change{\section{Dispersion Relations for Arbitrary Distributions}
\label{sec:code}}

\alps\ allows for the inclusion of any number of particle species or components with arbitrary gyrotropic background distributions $f_{0,s}(p_\perp,p_\parallel)$ as a function of momentum $\textbf{p}$, supporting normal modes with any direction of propagation, represented by the wavevector $\V{k} = \V{k}_\perp + \V{k}_\parallel$. 
Throughout this text, $\perp$ and $\parallel$ are defined with respect to the direction of the background magnetic field $\V{B}_0$.

Vlasov--Maxwell theory starts with the expression for the first-order perturbation of the current
\begin{equation}
    \mathbf{j}= \sum_s \mathbf{j}_s = \sum_s q_s \int d^3\mathbf{p}\, \mathbf{v} 
    \delta f_s(\mathbf{r},\mathbf{p},t)
    =    -\frac{i \omega}{4 \pi} \underline{\underline{\chi}}_s \cdot \mathbf{E},
    \label{eqn:perturbed_current}
\end{equation}
where $q_s$ is the charge of a particle of species $s$, $\omega$ is the complex wave frequency with real and imaginary components $\omega_{\textrm{r}}$ and $\gamma$, $\vec E$ is the electric field, $\underline{\underline{\chi}}_s$ is the susceptibility of species $s$, and $\delta f_s$ is the fluctuation of the VDF from the background $f_{0,s}$.
Following the careful application of identities, transformations, and substitutions outlined by \citet{Verscharen:2018} and covered in detail in Chapter 10 of \citet{Stix:1992}, we arrive at an expression for  $\underline{\underline{\chi}}_s$ in terms of integrals over functions of momentum derivatives of the background VDF
\begin{multline}
\label{eqn:chi_vm}
    \underline{\underline{\chi}}_s =  \frac{\omega_{p,s}^2}{\omega \Omega_{0,s}}    
        \int_0^\infty 2 \pi p_\perp d p_\perp \int_{-\infty}^\infty d p_\parallel
          \left[\hat{e}_\parallel \hat{e}_\parallel \frac{\Omega_s}{\omega} p_\parallel^2\right.\\
          \left.
         \times \left(\frac{1}{p_\parallel} \frac{\partial f_{s,0}}{\partial p_\parallel}
        -\frac{1}{p_\perp} \frac{\partial f_{s,0}}{\partial p_\perp}
         \right) + \sum_{n=-\infty}^{\infty} \frac{\Omega_s p_\perp U}{\omega - k_\parallel v_\parallel - n \Omega_s} \underline{\underline{T}}_n\right] 
\end{multline}
where
\begin{equation}
    U = \frac{\partial f_{0,s}}{\partial p_\perp} + \frac{k_\parallel}{\omega} 
    \left(v_\perp \frac{\partial f_{0,s}}{\partial p_\parallel}
    -v_\parallel \frac{\partial f_{0,s}}{\partial p_\perp}
    \right),
    \label{eqn:U}
\end{equation}
\begin{equation}
    \underline{\underline{T}}_n=
    \left(
  \begin{array}{ccc}
    \frac{n^2 J_n^2}{z^2} & \frac{i n J_n J_n'}{z} & \frac{n J_n^2 p_\parallel}{z p_\perp}\\
    -\frac{i n J_n J_n'}{z} & (J_n')^2 & \frac{-i J_n J_n' p_\parallel}{p_\perp}\\
     \frac{n J_n^2 p_\parallel}{z p_\perp}&  \frac{i J_n J_n' p_\parallel}{p_\perp}& \frac{J_n^2 p_\parallel^2}{p_\perp^2}\\
  \end{array}
    \right),
\end{equation}
and $J_n$ is the $n$th order Bessel function with argument $z = k_\perp v_\perp/\Omega_s$ and $J_n'$ is its derivative.
Characteristic timescales are defined in terms of the plasma frequency $\omega_{ps}=\sqrt{4 \pi n_s q_s^2/m_j}$ and the (signed) cyclotron frequency $\Omega_s = q_s B_0/m_s c$; $\Omega_{0,s}$ only includes the rest-mass in the denominator, while $\Omega_s$ includes the appropriate relativistic correction.

As described in App.~\ref{app:NHDS}, \alps\ may now use $ \underline{\underline{\chi}}_j$ from the bi-Maxwellian approximation as determined by \nhds\ \cite{Verscharen:2018a} or the cold plasma expression for any of the species' contribution to the dispersion relation.
This capability is useful when one of the components is well modeled as a fluid or the velocity distribution is not constrained by observations.

\alps\ has also been updated so that the user may specify any analytic form for $f_{0,s}(p_\perp,p_\parallel)$, allowing the direct calculation of the appropriate derivatives and integrals through quadrature on a specified grid, as well as the immediate extension to complex $p_\parallel$ values without the need for any fitting or polynomial representations during the evaluation of the analytic continuation.
While not applicable to cases where a simulation or model is not easily expressible in analytical form, it can be particularly useful when a theory predicts an analytical, but non-Maxwellian or non-$\kappa$ form (e.g., a bi-Moyal flattop distribution\cite{Klein:2016}) so that differences against standard bi-Maxwellian predictions can be quantified.

For the non-analytical contributions, \alps\ calculates $\partial f_{0,s}/\partial {p_\parallel}$ and $\partial f_{0,s}/\partial {p_\perp}$ on the prescribed grid and performs the necessary integrations, allowing the construction of the dielectric tensor
\begin{equation}
\underline{\underline{\epsilon}}(\omega,\mathbf{k})= \underline{\underline{1}}
+ \sum_s \underline{\underline{\chi}}_s(\omega,\mathbf{k})
\label{eqn:dielectric}
\end{equation}
which in turn allows the construction of the homogeneous linear wave equation
\begin{equation}
\underline{\underline{\Lambda}} \cdot \V{E} = 
\left(
\begin{array}{ccc}
    \epsilon_{xx} - n_z^2 & \epsilon_{xy} & \epsilon_{xz} + n_x n_z \\
   \epsilon_{yx} &  \epsilon_{yy} - n_x^2 - n_z^2& \epsilon_{yz} \\
    \epsilon_{zx }+ n_x n_z  &  \epsilon_{zy}  & \epsilon_{zz} - n_x^2 
\end{array}
\right) 
\left(
\begin{array}{c}
   E_x \\ E_y \\ E_z 
\end{array}
\right) =0
\label{eqn:wave_equation}
\end{equation}
where $\V{n}=c \V{k}/\omega$ is the complex index of refraction.
The solutions of $\Lambda=\det |\underline{\underline{\Lambda}}(\omega,\V{k})|=0$ are the normal modes supported by the prescribed background.
\alps\ identifies these solutions for fixed wavevectors as minima over a user-defined region of complex frequency space and follows identified solutions as a function of varying wavevector $\textbf{k}d_{\mathrm{ref}}$, where $d_{\mathrm{ref}} = c/\omega_{p,\mathrm{ref}}$ is the reference inertial length. 
In addition to determining dispersion relations, \alps\ also calculates the associated eigenfunctions of the fluctuating electromagnetic fields, densities, and velocities. These eigenfunctions represent the polarization of the identified plasma normal modes.
\alps\ also calculates the damping or growth rates associated with each species, and further separates the contributions to these rates from Landau, transit time, and cyclotron mechanisms. 
The details of the eigenfunction and heating calculations are included in App.~\ref{app:eigen}.

\change{\section{Improved Hybrid Analytic Continuation}
\label{sec:method}}

When the solutions are damped ($\gamma<0$, where $\gamma=\mathbb{I}\mathrm{m}(\omega)$) the integration of Eqn.~\ref{eqn:chi_vm} necessitates an analytic continuation of $f_{0,s}$ into the complex $p_\parallel$ plane.
This is straightforward if $f_{0,s}$ is a known analytical function which can be evaluated at a complex $p_\parallel$.\footnote{The relativistic case is more complex, given the non-trivial momentum dependence of the resonant denominator in that limit; see \S 3.3 by \citet{Verscharen:2018} for details, where we use the method suggested by \citet{Lerche:1967} to transform from $(p_\parallel,p_\perp)$ to $(\Gamma,\bar{p}_\parallel)$ and make the solution tractable.}
For the non-analytical case, as we only have values for $f_{0,s}$ for $\mathbf{p} \in \mathbb{R}$, we use two complementary schemes to extend the distribution function numerically.
The initial implementation of \alps\ is a fit of $f_{0,s}$ to a (typically small) number of analytical functions, \S~\ref{ssec:method.fit}.
In this work, we describe a newly implemented polynomial basis representation using a generalized linear least squares approach, \S~\ref{ssec:method.poly}. 
In both cases, we follow the prescription in \citet{Landau:1946} for the integration of a function $G$ over a contour $C_L$ that lies below the complex poles of the integrand:
\begin{align}
I(p_\perp) & = \int_{C_L} \mathrm{d}p_\parallel G(p_\perp, p_\parallel)
\label{eqn:landau}
\\ \nonumber
& = \begin{cases} 
\int_{-\infty}^{+\infty} \mathrm{d}p_\parallel G(p_\perp, p_\parallel) & \text{if } \gamma > 0, \\ 
\mathcal{P} \int_{-\infty}^{+\infty} \mathrm{d}p_\parallel G(p_\perp, p_\parallel) + i \pi \sum \mathrm{Res}_A(G) & \text{if } \gamma = 0, \\ 
\int_{-\infty}^{+\infty} \mathrm{d}p_\parallel G(p_\perp, p_\parallel) + 2 i \pi \sum \mathrm{Res}_A(G) & \text{if } \gamma < 0, 
\end{cases} 
\end{align}
where $\mathcal P$ indicates the principal-value integration, $\mathrm{Res}_A(G)$ is the residue of the function $G$ at point $A$, and the sum is taken over all poles $A$.
For the integral of interest in Eqn.~(\ref{eqn:chi_vm}), $G$ has one simple pole with
\begin{equation}
    \sum \mathrm{Res}_A(G) = - \frac{m_j}{|k_\parallel|} \Omega_j U \underline{\underline{T}}_n \bigg|_{p_\parallel = p_\text{pole}},
    \label{eqn:residual}
\end{equation}
which has to be evaluated for the six unique terms in $\underline{\underline{T}}_n$.

In the non-relativistic instance of the code, \alps\ evaluates the fit or representation separately at each value of $p_\perp$ on the grid, so that no assumption is made as to the structure of $f_{0,s}$ in the $p_\perp$-direction.
\alps\ uses these functions only if a pole is within the integration domain -- the momentum range provided in the input VDF grid -- and only if $\gamma \leq 0$. 
Therefore, the method of analytic continuation does not impact purely unstable solutions.

\subsection{Fit Function Representation}
\label{ssec:method.fit}

In the original version of \alps, in order to calculate the analytic continuation, the user specifies a small number of analytical functions (usually one or two, corresponding to core-and-beam structures inferred from in-situ solar wind observations \cite{Marsch:2012,Alterman:2018}), each with a handful of defining parameters.
The canonical case used is a linear combination of drifting bi-Maxwellians,
\begin{equation}
f_{0,s} = \frac{1}{\pi^{3/2} m_s^3 w_{\perp s}^2 w_{\parallel s}} \exp \left( - \frac{p_\perp^2}{m_s^2 w_{\perp s}^2} - \frac{(p_\parallel - m_s U_s)^2}{m_s^2 w_{\parallel s}^2} \right),
    \label{eqn:bimax}
\end{equation}
but bi-$\kappa$,
\begin{multline}
f_{0,s} =  \frac{1}{m_s^3 w_{\perp s}^2 w_{\parallel s}} \left[ \frac{2}{\pi (2 \kappa - 3)} \right]^{3/2} \frac{{\Gamma} (\kappa + 1)}{{\Gamma} (\kappa - 1/2)} \\
\times  \left\{ 1 + \frac{2}{2 \kappa - 3} \left[ \frac{p_\perp^2}{m_s^2 w_{\perp s}^2} + \frac{(p_\parallel - m_s U_s)^2}{m_s^2 w_{\parallel s}^2} \right] \right\}^{-(\kappa + 1)},
    \label{eqn:bikappa}
\end{multline}
where ${\Gamma}$ is the Gamma function, and J\"uttner distributions,
\begin{equation}
f_{0,s} = \frac{1}{2 \pi m_s^3 c w_s^2 K_2 (2 c^2 / w_s^2)} \exp \left( -2 \frac{c^2}{w_s^2} \sqrt{1 + \frac{|p|^2}{m_s^2 c^2}} \right),
    \label{eqn:juttner}
\end{equation}
are  included.
Additional functions, including two other representations of the J\"uttner distribution as well as a bi-Moyal distribution,
\begin{multline}
    f_{0,s}= A 
    \mathrm{exp}\left\{\frac{1}{2} \left[\frac{{p}^2_{\perp}}{m_s^2 w_{\perp s}^2} + \frac{({p}_{\parallel} -m_s U_s)^2}{m_s^2 w_{\parallel s}^2} \right.\right. \\
   \left.\left. -\mathrm{exp}(\frac{{p}^2_{\perp}}{m_s^2 w_{\perp s}^2} + \frac{({p}_{\parallel}-m_s U_s)^2}{m_s^2 w_{\parallel s}^2}) \right]\right\},
    \label{eqn:bimoyal}
\end{multline}
with $A$ a normalization constant, have been added since the initial code release.

At each $p_\perp$ value, a 1-D Levenberg--Marquardt fit\citep{Levenberg:1944,Marquardt:1963} is performed as a function of $p_\parallel$.
For $p_\perp=0$,  user-defined inputs are used to initialize the fit parameters.
\alps \ uses the final fit results from $p_\perp=0$ as the initial guess for the next $p_\perp$ value, eventually iterating across the entire $p_\perp$ range.
This process works well for sufficiently smooth functions and fit initializations sufficiently close to a minimum for the Levenberg--Marquardt algorithm.

However, when applying the code to more complicated VDFs, two recurrent difficulties can arise \change{that necessitate the development of a new representation method}.
First, even when a good set of fit initializations were selected, the fitting routine could fail at larger $p_\perp$ values, typically due to pathological but not unphysical variations of the VDF as a function of $p_\perp$, e.g. the disappearance of a secondary population or the appearance of an additional population at larger $p_\perp$.
This failure would cause the entire fitting procedure to halt.
Second, even when the fits converge for all $p_\perp$, the small number of functions used for the analytical representation lead to inaccuracies in the representation of the VDF, producing discrepancies as dispersion relations cross from $\gamma>0$ to $\gamma<0$ or vice versa. 

These discrepancies arise for two reasons.
First, the momentum derivatives of $f_{0,s}$ and the fit function can differ when evaluated at the resonant velocity $v_{\textrm{res}}=(\omega-n \Omega_s)/k_\parallel$. 
Secondly, the total velocity moments of the fit function can differ from the velocity moments of the actual VDF, leading to an effective net charge and current that is different for $\gamma>0$ and $\gamma<0$.
In addition to producing incorrect mode structure for damped solutions, these discrepancies can lead to a sharp edge at $\gamma=0$ that causes the numerical Newton-secant search used to follow dispersion relations with varying wavevector to jump to a different solution or lose the physical solutions entirely due to the large \change{unphysical} gradient in $\Lambda$.
Such a jump is illustrated in the surfaces associated with low-order fits in Fig.~\ref{fig:method_compare}, e.g. the grey line in the lower right panel.

These issues \change{with the fit function representation for VDFs that are significantly different from a finite superposition of Maxwellian functions in the $p_{\parallel}$ dimension} have led us to the development of a new approach to perform the hybrid analytic continuation, \change{combining numerical integration of the input VDF with a \textit{polynomial basis representation} when the dispersion relation requires evaluation of the VDF at complex momentum values.}

\subsection{Polynomial Basis Representation}
\label{ssec:method.poly}

In order to reduce the discrepancies associated with representing $f_{0,s}$ using one or a handful of physically representative fitted analytical functions, we have implemented a Generalized Linear Least Squares  (GLLS)representation of $f_{0,s}(p_\parallel)$ for fixed values of $p_\perp$, following \S 15.4 of \citet{Press:1992}.
Specifically, we find the parameters $a_k$ for the model
\begin{equation}
    y(x)=\sum_{k=0}^{M-1} a_k X_k(x),
\end{equation}
where $X_k(x)$ are a set of basis functions and $y(x)$ is the data to be fit.
For this implementation, we have selected the Chebyshev polynomials $T_k(x)$ of the first kind  as the basis functions, which are defined recursively through
\begin{align}
T_0(x)& =1,\\
T_1(x)&=x,\\
T_{n+1}(x)&=2x T_n(x)-T_{n-1}(x).
    \label{eqn:Tn}
\end{align}
As $T_n$ are bound between $[-1,1]$, we perform a change of variables from $p_\parallel$ to $x$ through
\begin{equation}
    x=\frac{p_\parallel -\left(p_\parallel^{\text{max}}+p_\parallel^{\text{min}} \right)/2}{\left(p_\parallel^{\text{max}}-p_\parallel^{\text{min}} \right)/2}
\end{equation}
where $p_\parallel^{\text{max,min}}$ are the maximum and minimum values of the input VDF grid.

The bounded nature of $T_n$ is ideally suited to the limited range of $p_\parallel$ used as inputs for \alps.
Chebyshev polynomial representation performs better than other classes of orthogonal polynomials, e.g. the Hermite or Legendre polynomials, not shown.
We model $\log_{10}[f_{0,s}]$ rather than $f_{0,s}$ to capture the full range of the VDF amplitude and not solely the structure near the peak of the distribution.
We emphasize that, as with the functional fitting approach discussed in \S~\ref{ssec:method.fit}, this is a purely one-dimensional method, representing $f_{0,s}(p_\parallel)$ separately for the $n_\perp$ rows in the input VDF grid; we write the VDF slice that we represent as $y_i$, which is known at $N$ values of $x$.

The solution is performed via the Normal Equations method, solving
\begin{equation}
    \sum_{j=0}^{M-1}\alpha_{kj}a_j = \beta_k,
    \label{eqn:normal}
\end{equation}
where 
\begin{equation}
    \alpha_{kj}= \sum_{i=0}^{N-1} \frac{X_j(x_i)X_k(x_i)}{\sigma_i^2}
\end{equation}
and 
\begin{equation}
    \beta_k=\sum_{i=0}^{N-1}\frac{y_i X_k(x_i)}{\sigma_i^2}.
\end{equation}
If the errors in the knowledge of $y_i$ are known, one can set $\sigma_i$ to non-unity values; in the current implementation, $\sigma_i = 1 \forall x_i$. 
Eqn.~\ref{eqn:normal} is solved using the \texttt{dgemm}, \texttt{dgemv}, and \texttt{dgesv} routines from the \texttt{LAPACK} library. 
\change{As the GLLS solution is calculated once for an input VDF, there is no need to implement more efficient existing schemes for determining $a_{kj}$ that exploit connections between Fourier and Chebyshev representations. \citep[e.g.][]{Mason:2002}}

\begin{figure*}
    \includegraphics[width=0.95\columnwidth]{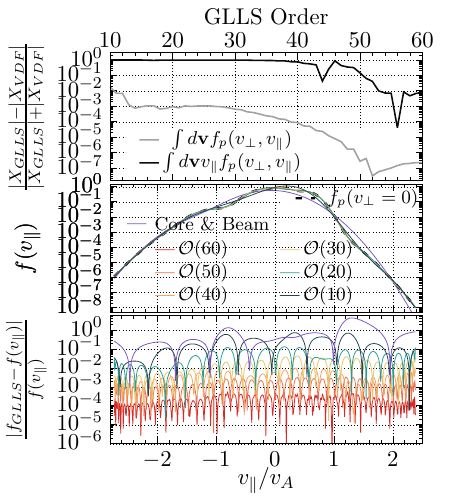}
        \includegraphics[width=1.05\columnwidth]{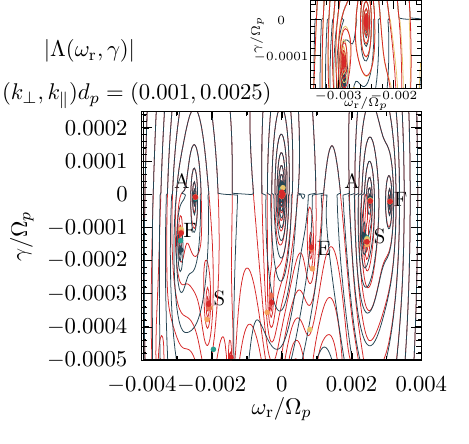}
    \caption{(Left) Comparison of input $f_{0,p}(v_\perp=0,v_\parallel)$ with GLLS Chebyshev representations and a fit consisting of the sum over two bi-Maxwellians. (Top) relative percentage difference in the density (grey) and current (black) between the GLLS and input VDFs as functions of the GLLS order. The middle and bottom panels show the reconstructed $f_{0,p}$ at $v_\perp=0$ as well as the relative error compared to the input VDF. 
    (Right) Isocontours of constant $\Lambda(\omega_{\textrm{r}},\gamma)$ for \alps\ solutions using the GLLS analytic continuation for $\mathcal{O}(10)$ (grey dashed) and $\mathcal{O}(60)$ (red), using $(k_\perp,k_\parallel)d_p = (0.001,0.0025)$.
    The above inset illustrates the transition between the extreme orders used in this work over a limited region of complex frequency space.
    Identified solutions with $\Lambda=0$ are shown with colored dots using the same color scale as the left panels, with letters indicating Slow, Alfv\'en, Fast, or Entropy mode (SAFE) solutions.
    }    
    \label{fig:method_compare}
\end{figure*}

The polynomial basis method produces superior results compared to the functional fitting method for highly structured VDFs, both in terms of the smoothness of $\Lambda(\omega)$ across $\gamma=0$ due to improved accuracy of $\partial_p f_{0,j}$ for all $\omega$ and $\V{k}$ values, as well as in correctly capturing the total density and current of the VDFs.
To illustrate this, Fig.~\ref{fig:method_compare} shows the representation of a slice of $f_{0,p}(p_\parallel)$ for $p_\perp=0$ using the GLLS method for orders $N=(10,20,30,40,50,60)$ and the best fit to the sum of two bi-Maxwellians.
Also shown are the relative difference between the fit or GLLS representation and the actual VDF slide.
We consider a distribution similar to one from \citet{Walters:2023}, interpolated from Wind SWE observations to a fixed grid of $300 \times 600$ points in $(p_\perp \times p_\parallel)$.
Looking at the $p_\perp=0$ slice, neither the bi-Maxwellian fit (purple) nor the lowest order GLLS representation (grey) of $f_p(p_\parallel)$ match the actual distribution, with the relative error approaching or exceeding order unity.
Moving to higher order representations, the relative error decreases by several orders of magnitude; the highest order we consider, $\mathcal{O}(60)$, is well below the number of resolved points in $p_\parallel$, ensuring we are not over-fitting the distribution.

In the right-hand panels of Fig.~\ref{fig:method_compare}, we plot isocontours of the magnitude of  $\log_{10}[\Lambda(\omega_{\textrm{r}},\gamma)]$, as well as the associated solutions satisfying $\Lambda=0$, for a wavevector of $(k_\perp,k_\parallel)d_p = (0.001,0.0025)$.
The isocontours for both the lowest, $\mathcal{O}(10)$, and highest, $\mathcal{O}(60)$, representations are shown.
\change{For the $\mathcal{O}(10)$ case (grey dashed lines), there is a stark discontinuity across $\gamma=0$, indicating that the VDF representation is not sufficiently accurate in its representation of the bulk velocity moments and resonant momentum gradients involved in the principal-value integration, Eqn.~\ref{eqn:landau}, to enable a smooth transition between growing and damping solutions.}
For the $\mathcal{O}(60)$ case (red), the isocontours are continuous across $\gamma=0$; the transition between these two orders over a limited region of complex frequency is illustrated in the upper right inset, illustrating the convergence towards a smooth surface with higher polynomial order.

In addition to the $\gamma=0$ discontinuity, the damping rates of the moderately damped modes are significantly altered depending on the order of the GLLS representation.
The backward-propagating Fast solution -- with the sign of $\omega_{\textrm{r}}$ indicating the direction of propagation -- becomes significantly less damped with increasing order, while the forward-propagating Slow solution becomes more damped.
We see the complex frequencies of the moderately damped solutions begin to converge at $\mathcal{O}(50)$, which defines the order that we use for the remainder of this work.

In considering a wider set of test VDFs, we find that the error in the density (zeroth velocity moment of $f_{0,s}$) and background current (first velocity moment of $f_{0,s}$) of the reconstructed VDF serves as a good proxy for deciding whether the reconstruction order is sufficiently high for our hybrid analytic continuation method to work.
The relative error between the representation and the input  VDFs is shown for both quantities as a function of order in the top panel of Fig.~\ref{fig:method_compare}. 
We see that good agreement can be achieved with relatively low order for the density, but that higher orders are needed to reduce the error in the current.
We find that a relative percent difference\footnote{We use relative percent difference rather than relative error as we are frequently evaluating the dispersion relation in the frame where the parallel velocity moment of the VDF under consideration is effectively zero.} $|X_{GLLS}-X_{VDF}|/|X_{GLLS}+X_{VDF}|$ of less than $0.1$ is typically sufficient for the method to yield reliable results.

\begin{figure}
    \centering
    \includegraphics[width=0.95\linewidth]{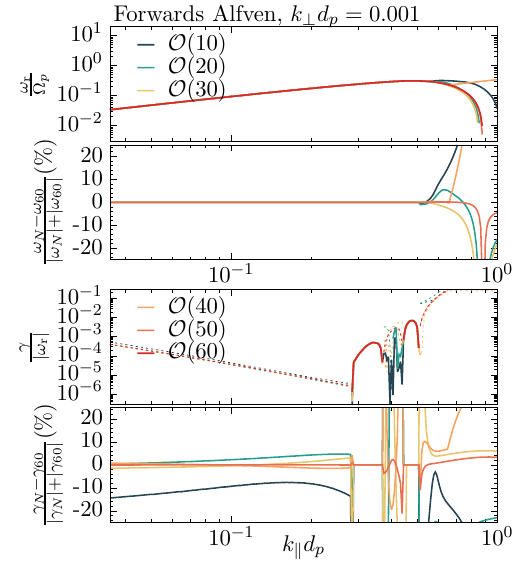}
    \caption{Real and imaginary parts of the complex frequency $\omega = \omega_{\textrm{r}} + i \gamma$ for the parallel propagating Alfv\'en/ion cyclotron wave using GLLS representations with $\mathcal{O}(10)$ through $\mathcal{O}(60)$.
    The relative error between the lower-order and $\mathcal{O}(60)$ solutions illustrates the order necessary for convergence for the input VDF.}
    \label{fig:order-dispersion}
\end{figure}

While a relative error of the current of less than $10\%$ is generally sufficient for the solutions to converge, for the case of more pathological VDFs, additional convergence studies should be considered.
An example convergence study for the same VDF as used in Fig.~\ref{fig:method_compare} is shown in Fig.~\ref{fig:order-dispersion} for the forwards propagating Alfv\'en solution varying $k_\parallel d_p$ for fixed $k_\perp d_p = 0.001$.
Both the real and imaginary parts of the complex frequency $\omega$ are shown, alongside the relative error compared to the solution based on GLLS of order $\mathcal{O}(60)$.
We see that $\omega_{\textrm{r}}$ is largely unaffected until significant damping arises, at which point the differences in the damping rates change the speed of propagation of the modes.
The growth rates  $\gamma >0$ are correct for all orders as expected due to the analytic continuation not being invoked for these solutions according \change{to Eq.~(\ref{eqn:landau})}.
The damped modes have $5-10\%$ differences for the low-order solutions compared to the converged solution.
Moreover, when transitioning across $\gamma=0$, the lower-order representations can jump to unphysical solutions.
The impact of representations with unnecessarily high order  on strongly damped solutions are discussed in App.~\ref{app:cheb}.

\section{Comparing Dispersion Relations}
\label{sec:compare}

\subsection{Validating against bi-Maxwellian Models}
\label{ssec:2-vdfs}

\begin{figure}
    \centering
    \includegraphics[width=0.95\linewidth, trim=0pt 10pt 290pt 15pt, clip]{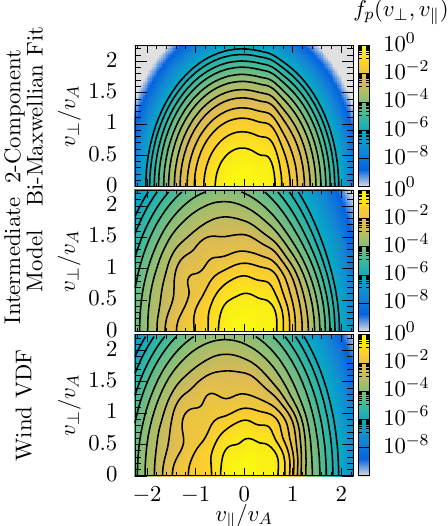}
    \caption{{\color{red}}
    Input proton VDFs used for the \alps\ calculations, including the best-fit two-bi-Maxwellian $f_{C\&B}$ (top), observed Wind VDF $f_{Obs.}$ (bottom),  and one intermediate model, Eqn.~\ref{eqn:intermediate} (middle).
    }
    \label{fig:vdfs}
\end{figure}

\begin{figure}
    \centering
            \hspace*{-1.35cm}
    \includegraphics[width=0.665\columnwidth, trim=0pt 12pt 0pt 0pt, clip]{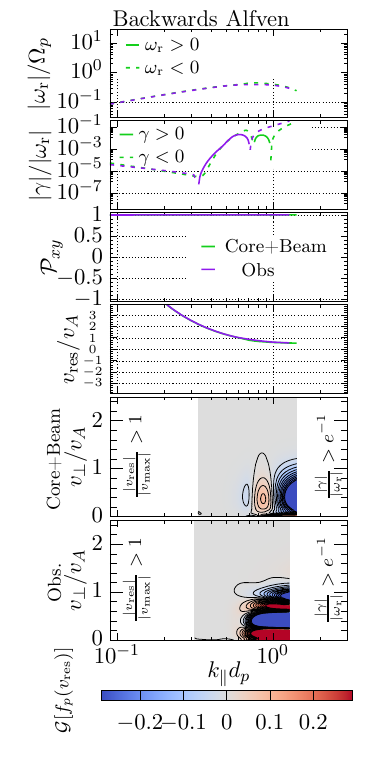}
        \hspace*{-1.35cm}
    \includegraphics[width=0.61\columnwidth, trim=15pt 12pt 0pt 0pt, clip]{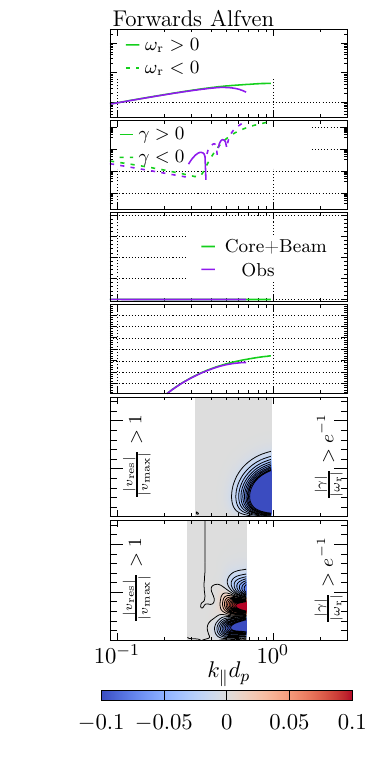}
    \caption{Dispersion relations for parallel-propagating \Alfven waves supported by the observed (green lines) and core-and-beam (purple) VDFs:
      (top row) real part of the frequency $\omega_{\textrm{r}}/\Omega_p$,
      (second) growth (solid) and damping (dashed) rates per wave period $\gamma/\omega_{\textrm{r}}$,
      (third) polarization $\mathcal{P}_{xy}$, and
      (fourth) resonant velocity $v_{\textrm{res}}/v_A$.
      The bottom two rows show the quasilinear operator applied to the resonant velocity associated with wavevector, $\mathcal{G}\{f_p[v_\parallel^{\textrm{res}}(k_\parallel d_p)]\}$ from Eqn.~\ref{eqn:Gfv}.
      The structure in this space illustrates the velocity regions responsible for wave emission (red) and absorption (blue).}
    \label{fig:QL-Wind}
\end{figure}

We next turn to the consideration of the relation between the \alps\ solutions and those calculated based on multi-component bi-Maxwellian fits to the background VDF.
To investigate this, we consider the same observed proton VDF measured by  the Wind spacecraft of the solar wind interval considered in \S~\ref{sec:method} -- $f_{Obs}(v_\perp,v_\parallel)$ -- as well as a two-component bi-Maxwellian best fit of the proton VDF, $f_{C\&B}(v_\perp,v_\parallel)$ imposed on a discrete velocity grid.
\change{These VDFs are shown in the top and bottom panels of Fig.~\ref{fig:vdfs}.}
\footnote{\change{As we are focused on modeling thermal populations in the solar wind, for the remainder of the manuscript we consider $f_{0,s}$ as a function of velocity rather than momentum.}}
We use these two input VDFs to calculate the associated linear dispersion relations, shown in Fig.~\ref{fig:QL-Wind} for both the backwards and forwards parallel-propagating \Alfven solutions;
$k_\perp d_p$ is fixed at $0.001$.
Both VDFs have the same velocity resolution, and the \alps bi-Maxwellian solution agrees with a `traditional'  calculation from \plume\ (not shown).
As previously noted by \citet{Walters:2023}, the real parts of the frequency $\omega_{\textrm{r}}/\Omega_p$ for the solutions based on $f_{Obs}$ and $f_{C\&B}$ are qualitatively the same at large scales, deviating only when the damping is sufficiently strong to slow the propagation of the wave.
For clarity, we only plot solutions for which $|\gamma|/|\omega_{\textrm{r}}| < 1/e$, indicating they are not too strongly damped.

The wavevector regions of instability are significantly different when using $f_{Obs}$ or $f_{C\&B}$.
The forward \Alfven mode is stable for the core-and-beam VDF, but supports two regions of instability, with the smaller-scale region having a moderate growth rate.
Both backwards \Alfven modes support instabilities, but the observed VDF is more unstable, with a broad range of unstable wavevectors.

The perpendicular electric field polarization of the solutions\cite{Verscharen:2013c}
\begin{equation}
\mathcal{P}_{xy}=
\frac{|E_{R}|-|E_{L}|}{|E_{R}|+|E_{L}|}
\ni E_{R,L}\equiv \frac{E_x \mp iE_y}{\sqrt{2}}
    \label{eqn:polarization}
\end{equation}
as well as the resonant velocities
\begin{equation}
v_\parallel^{\textrm{res}}=\frac{\omega- n \Omega_p}{k_\parallel},
\end{equation}
where the left handed waves (forward \Alfven and backward fast) have $n=1$ and the right handed waves (backwards \Alfven and forward fast) have $n=-1$, 
are effectively identical between the solutions from the two models.

What drives these differences can be identified using quasilinear theory\cite{Kennel:1966a,Verscharen:2013c}.
The resonant coupling of the wave to the VDF can be represented in terms of the operator
\begin{equation}
\mathcal{G} \equiv \left(1-\frac{k_\parallel v_\parallel}{\omega_{\textrm{r}}} \right)
\frac{\partial}{\partial v_\perp}+
\frac{k_\parallel v_\perp}{\omega_{\textrm{r}}}
\frac{\partial}{\partial v_\parallel}
\label{eqn:ql}
\end{equation}
\change{applied to $f_p(v_\perp,v_\parallel)$.}
To illustrate the structures of the VDF responsible for growth and damping, we construct a $k_\parallel$-dependent function, applying $\mathcal{G}$ to $f_p$ at $v_\perp$ and the resonant parallel velocity associated with each wavevector,
\begin{equation}  \mathcal{G}\left[f_p\left(\frac{v_\parallel^{\textrm{res}}(k_\parallel d_p)}{v_A},\frac{v_\perp}{v_A}\right)\right] \ni \frac{v_\parallel^{\textrm{res}}(k_\parallel d_p)}{v_A}=\frac{\omega_{\textrm{r}}/\Omega_p-n}{k_\parallel d_p}.
  \label{eqn:Gfv}
\end{equation}
Velocities are normalized to the \Alfven velocity $v_A=B/\sqrt{4 \pi n_p m_p}$.
This function is illustrated in the bottom two rows of Fig.~\ref{fig:QL-Wind}.
As this calculation is only valid \change{in the weak damping or growth limit\cite{Kennel:1966a} $|\gamma|/|\omega_{\textrm{r}}|<1/e$,} and only useful when the resonant velocity samples the input VDF, we restrict ourselves to the limited range or wavevectors satisfying both these conditions.
The sign of this function at each velocity value indicates whether that point in phase space emits or absorbs power in interaction with the given wave.
As illustrated by \citet{Walters:2023}, the sign of the function integrated over $v_\perp$ indicates whether the species has a net absorption or emission at that parallel scale.

For the forwards \Alfven solution, there is no portion of the core-and-beam VDF that drives an instability.
For the observed VDF, structures at $v_\parallel \sim - v_A$ and $v_\perp \in [0.5,1] v_A$ are responsible for the range of strongly unstable modes; isocontours of the VDFs are shown in the left-hand column of Fig.~\ref{fig:alfven-example}.
For the backwards \Alfven waves, the structure responsible for emission is starkly different between the two VDF representations, with all $v_\perp$ values coherently contributing to the core-and-beam emission over a narrow band of $v_\parallel$ values. 
The observed VDF has a striated structure in $v_\perp$ with different velocity regions competing to enhance or reduce the wave emission at each $k_\parallel$.

\begin{figure}
    \centering
    \includegraphics[width=0.95\linewidth, trim=35pt 5pt 0pt 0pt, clip]{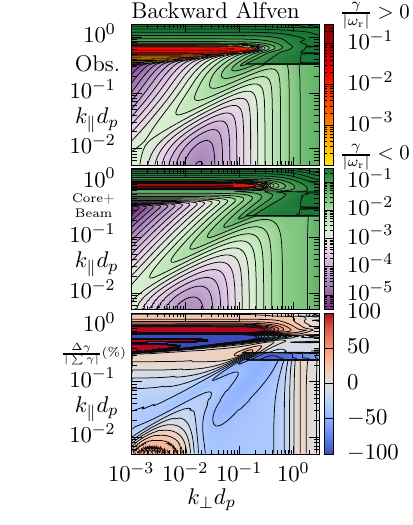}
    \caption{Damping and growth rates for $f_{Obs}$ (top) and $f_{C\&B}$ (middle) for the backwards propagating \Alfven wave solution. 
    The percent difference between the two models is shown in the bottom panel, highlighting significant reductions in the kinetic \Alfven wave damping rates and increase in the ion cyclotron wave emission rates for the observed VDF compared to the bi-Maxwellian representation.
    }
    \label{fig:kperp-kpar-damping}
\end{figure}

We can further characterize differences in the damping or emission as a function of scale, shown in Fig.~\ref{fig:kperp-kpar-damping}.
While qualitatively having similar structure, namely increased resonant coupling as scales reach $|k|d_p \sim 1$, the details of the VDF lead to significant differences in the predicted wave emission and absorption, quantified as $\frac{\gamma_{Obs}-\gamma_{C\&B}}{|\gamma_{Obs}+\gamma_{C\&B}|}$.
The differences in the parallel ion cyclotron branch agree with the parallel cut illustrated in Fig.~\ref{fig:QL-Wind}.
For $k_\perp d_p \lesssim 1$, the \Alfven wave transitions to a kinetic \Alfven wave, becoming strongly damped for both cases.
However, the observed VDF has a significantly reduced damping rate compared to the core-and-beam response; a similar result is found for the forwards solution.
Such changes in the damping rates may impact the overall heating rates used in a variety of models for hot, weakly collisional systems\cite{Howes:2010b,Chael:2018,Gorman:2024,Shankarappa:2024}, that use bi-Maxwellian models for the determination of $\gamma$ to inform the bifurcation of energy between ions and electrons.

\section{Continuity of Solutions}
\label{ssec:continuity}

\begin{figure}
    \centering
    \includegraphics[width=1.\columnwidth, trim=5pt 15pt 0pt 0pt, clip]{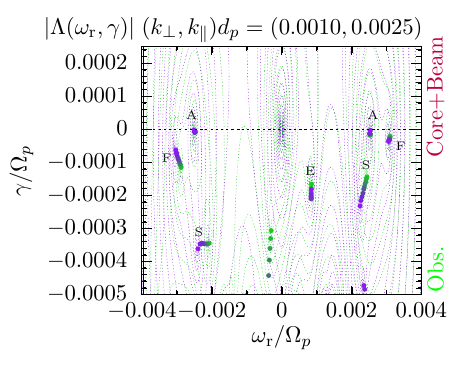}
    \caption{Isocontours for constant values of the dispersion surface $\Lambda(\omega_{\textrm{r}},\gamma)$ for \alps\ solutions using $f_{Obs.}$ (green lines) and $f_{C\&B}$ (purple) at constant wavevector $(k_\perp,k_\parallel)d_p=(0.001,0.0025)$.
      Identified Slow, Alfv\'en, Fast and Entropy solutions satisfying $\Lambda=0$ are shown with dots.
      The continuity between the $f_{Obs.}$ and $f_{C\&B}$ solutions are shown with the intermediate $f_i$ solutions, indicated by color.
      }
    \label{fig:map_explore}
\end{figure}

To characterize the continuity between the observed and core-and-beam solutions, we determine dispersion relations associated with an ensemble of intermediate VDFs between the two models.
At each point of the velocity grid, we calculate $\Delta f \equiv f_{Obs.}-f_{C\&B}$, and then construct $N$ VDFs that continuously vary between the models as
\begin{equation}
    f_{i}(v_\perp,v_\parallel)= f_{Obs.}(v_\perp,v_\parallel)+
    \frac{i \Delta f(v_\perp,v_\parallel) }{N-1}
    \label{eqn:intermediate}
\end{equation}
for $i \in [1, N]$. We then evaluate the \alps\ solutions for all $f_i$ and show the transition of the solutions between models based on $f_{Obs.}$  and $f_{C\&B}$.
For this work, $N=10$ elements are sufficient to smoothly track the change in the mode characteristics.

The normal mode structure in complex frequency space for the solutions associated with the full ensemble of VDFs at a fixed $(k_\perp,k_\parallel)d_p=(0.001,0.0025)$ is illustrated in Fig.~\ref{fig:map_explore}.
The solutions from the intermediate VDFs produce a continuous path between the solutions based on the core-and-beam fit (purple) and the observed representation (green), indicating that \alps\ produces solutions that smoothly vary between the models.
At the selected wavevector, $\omega_{\text{r}}$ remains relatively unchanged for all $f_i$.
For the damping rates, some of the solutions are qualitatively unaffected at this wavevector, e.g. both Alfv\'en solutions and the forward fast mode, by the variation in the underlying VDF.
The compressive slow and entropy modes, as well as the backwards fast mode, undergo significant changes in their damping rates when transitioning from $f_{Obs.}$  to $f_{C\&B}$.
There is not a universal increase or decrease in the damping rates when transitioning between the two models, but rather each mode responds uniquely to the variation, e.g., the backwards fast mode becoming more damped and the forwards slow mode becoming less damped for the observed VDF.

\begin{figure}
    \centering
    \includegraphics[width=0.975\linewidth, trim=216pt 7pt 15pt 15pt, clip]{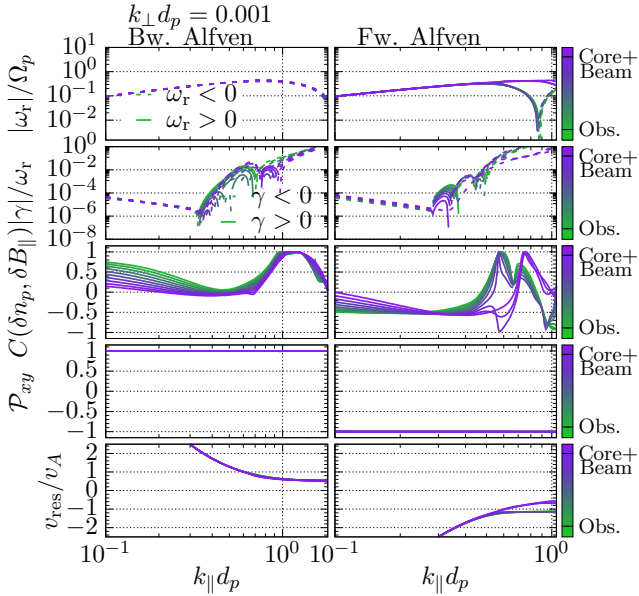}
    \caption{{\color{red}}
    Real part of the frequency $\omega_{\textrm{r}}/\Omega_p$, damping (dashed lines) and growth (solid) rates $\gamma/\omega_{\textrm{r}}$, the density-$B_z$ correlation, electric field polarization $\mathbf{P}_{xy}$, and $v_{\textrm{res}}$ as a function of $k_\parallel d_p$ for fixed $k_\perp d_p$ for the forward and backward \Alfven solutions using the continuous set of VDFs.
    }
    \label{fig:alfven-example}
\end{figure}

We next turn to a consideration to the variations of  these solutions with wavevector, in particular determining the impacts of the observed distribution on the wave's  eigenfunction characteristics.
In Fig.~\ref{fig:alfven-example}, we plot the real and imaginary parts of $\omega$ for the backwards and forwards \Alfven waves.
The strong damping due to the proton cyclotron resonance acts to slow, and eventually reverse, the propagation direction of the forwards \Alfven solutions near $k_\parallel d_p \sim 1$.
For the backwards solution, $\omega_{\textrm{r}}$ is qualitatively the same across $k_\parallel$ for all VDFs $f_i$ from the ensemble.
The perpendicular electric field polarization $\mathcal{P}_{xy}$ of the waves is constant across all evaluations, but the correlation between the density and parallel magnetic field fluctuations
\begin{equation}
    C(\delta n_p, \delta B_\parallel) = Re\left(\frac{\delta n_p \delta B_\parallel^*}{|\delta n_p| |\delta B_\parallel|}\right)
\end{equation}
is altered, with the amplitude of $C(\delta n_p,\delta B_\parallel)|$ increasing at large scales for the more observation-like VDFs (i.e., for low $i$).
Given that such eigenfunction relations are frequently invoked as a means of identifying wave modes in space plasmas \cite{Krauss-Varban:1994,Klein:2012,Verscharen:2017}, these changes in the eigenfunctions impact some interpretations of what waves are present in magnetospheric and solar wind plasmas.

\begin{figure*}
    \centering
    \includegraphics[width=1.0\linewidth,trim=27pt 0pt 0pt 0pt, clip]{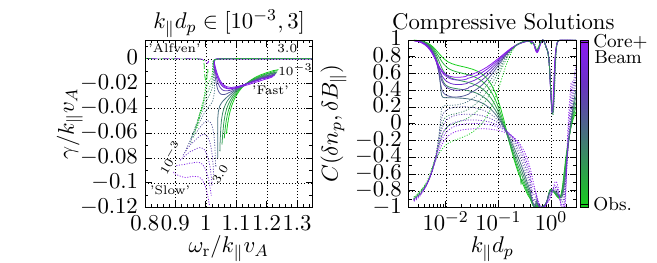}
    \caption{Dispersion relations for the forward propagating fast and slow magnetoacoustic modes for the core-and-beam VDF $f_{C\&B}$, the observed VDF $f_{Obs.}$, and intermediate VDF representations $f_i$.
    (left panel) 
    Parametric curves of $[\omega_{\text{r}},\gamma](k_\parallel,d_p)$ show distinct connections between the MHD-scale and ion-scale solutions for the observed and core-and-beam models.
    (right) Variation in $C(\delta n_p, \delta B_\parallel)$ of the fast and slow solutions as a function of parallel wavevector.}
    \label{fig:omega-gamma-rep}
\end{figure*}

The behavior of the fast and slow magnetosonic modes is also impacted by the velocity space structure.
In Fig.~\ref{fig:omega-gamma-rep}, we plot the dispersion relations for the forward parallel-propagating fast and slow solutions for the observed, core-and-beam, and intermediate VDFs.
The forward fast and slow modes undergo a mode conversion transitioning between the observed and fit distributions, similar to those seen near exceptional points \footnote{Exceptional points satisfy $\Lambda(\omega)=0$ and $d_\omega \Lambda(\omega)=0$ and act as branch points that allow continuous variations in the inputs to the dispersion relation to smoothly vary between different normal mode solutions; c.f. the Appendix of \citet{Klein:2015a}}. 
The solution that is more weakly damped at MHD scales has a greater phase speed and a positive correlation between $\delta n_p$ and $\delta B_\parallel$ when using the core-and-beam VDF. 
It remains the faster, more weakly damped, and positive-$C$ solution at smaller parallel wavevectors. 
The slower mode at MHD scales similarly remains more strongly damped and negatively correlated in $C$ as the solution approaches $k_\parallel d_p \sim 1$.

However, when using the observed VDF, the faster and slower solutions at large scales exchange their phase speeds and senses of correlation between $\delta n_p$ and $\delta B_\parallel$ at smaller scales, with the `slow' MHD mode becoming faster than the `fast' MHD mode.
This change in the mode behavior is driven by a change in the correlation between the density and magnetic field fluctuations, shown in the right panel of Fig.~\ref{fig:omega-gamma-rep}.
When using the core-and-beam model, the fast mode phase speed is enhanced by the in-phase response of $\delta n_p$ and $\delta B_\parallel$, while the slow mode phase speed is lessened by its out of phase response.
When using the observed VDF, this phase changes for the two modes around $k_\parallel d_p \sim 0.1$.
By investigating the phases for the models using the intermediate VDF representations, we see that this exchange is not a numerical artifact, but is a smooth transition between the calculations with $f_{Obs.}$ and $f_{C\&B}$.
Such variations in the correlation in plasmas with non-bi-Maxwellian background VDFs may impact the characterization of mode composition\cite{Klein:2012,Verscharen:2017}, as well as the nonlinear coupling and participation in the turbulence cascade of compressible solutions \cite{Schekochihin:2009,Kawazura:2020} and warrants further study.


\section{Conclusions}
\label{sec:conclusions}

We have extended the Arbitrary Linear Plasma Solver (\alps) by introducing a robust hybrid analytic continuation framework using high‐order polynomial (Chebyshev) representations of arbitrary velocity distribution functions (VDFs) to determine the associated dielectric plasma response.
The new Chebyshev generalized linear least squares (GLLS) approach delivers smooth and accurate continuations of the dispersion relation solutions to the linear Vlasov--Mawell equations across $\gamma=0$, eliminating discontinuities associated with low‐order analytical fits and enabling reliable tracking of normal modes from unstable to damped regimes.

Through systematic comparisons between spacecraft‐observed solar‐wind proton VDFs and their best‐fit two‐component bi‐Maxwellian representations, we demonstrate that the updated \alps\ produces continuous solutions to the plasma dispersion relation between bi‐Maxwellian and arbitrary VDF inputs.
We find that mode polarizations and heating rates can vary significantly when accounting for non-(bi‐)Maxwellian structure in the VDF, impacting interpretations of \Alfven, fast, and slow modes.

These advancements extend \alps’s applicability to a broad range of plasma environments such as the solar wind, magnetosheath, and astrophysical systems, where deviations from Maxwellian equilibria govern wave behavior and heating processes. 
By coupling numerical integration for arbitrary components with efficient bi‐Maxwellian and cold‐plasma models for fluid‐like species, the updated hybrid solver offers improvements in  computational speed. 
We anticipate that the improved analytic continuation and eigenfunction outputs will enable improved theoretical predictions and guide the interpretation of in-situ spacecraft measurements in future studies.

\section*{Acknowledgments}

The authors thank the UCL Open Source Software Sustainability Project for their assistance in publishing \alps\ as an open source code.
This study benefited from support by the International Space Science Institute (ISSI) in Bern, through ISSI International Team project 24-612 ("Excitation and dissipation of kinetic-scale fluctuations in space plasmas").
K.G.K was supported in part by grant no. NSF PHY-2309135 to the Kavli Institute for Theoretical Physics (KITP) and NASA grant No. 80NSSC19K0912 and 80NSSC24K0171. 
D.V.~is supported by STFC Consolidated Grant ST/W001004/1.

\section*{Data Availability Statement}

The \alps\ code is available via an open source BSD 2-Clause License at \url{https://github.com/danielver02/ALPS} with a full tutorial on its use at \url{https://danielver02.github.io/ALPS/} \citep{ALPS:2023}.

\appendix

\section{Using bi-Maxwellian and Cold Plasma Models for Select Species}
\label{app:NHDS}

For numerical expediency, \alps\ been updated to allow users to declare the susceptibility for any given component $\underline{\underline{\chi}}_s$ to be calculated using the numerical integration or using a bi-Maxwellian or cold plasma model. 
In all cases, \alps\ solves Eq.~(\ref{eqn:wave_equation}), but with different representations of $\underline{\underline{\chi}}_s$.
For the bi-Maxwellian case, we use the susceptibility as evaluated from \nhds \citep{Verscharen:2018a}, which follows the notation by \citet{Stix:1992}. 
The bi-Maxwellian susceptibility is given by
\begin{equation}
\underline{\underline{\chi}}_s=\hat{\vec e}_{\parallel}\hat{\vec e}_{\parallel}\frac{2\omega_{p,s}^2U_s}{\omega k_{\parallel}w_{\perp s}^2}+\frac{\omega_{p,s}^2}{\omega}\sum\limits_{n=-\infty}^{\infty}e^{-\lambda}\underline{\underline{Y}}_{n,s},
\end{equation}
where
\begin{equation}
\underline{\underline{Y}}_{n,s}=\begin{pmatrix}\frac{n^2I_n}{\lambda}A_n & -in(I_n-I_n')A_n & \frac{k_{\perp}}{\Omega_s}\frac{nI_n}{\lambda}B_n \\
in(I_n-I_n')A_n & \left(\frac{n^2}{\lambda}I_n+2\lambda I_n-2\lambda I_n'\right)A_n & \frac{ik_{\perp}}{\Omega_s}(I_n-I_n')B_n \\
\frac{k_{\perp}}{\Omega_j}\frac{nI_n}{\lambda}B_n & -\frac{ik_{\perp}}{\Omega_s}(I_n-I_n')B_n & \frac{2(\omega-n\Omega_s)}{k_{\parallel}w_{\perp s}^2}I_nB_n\end{pmatrix},
\end{equation}
$I_n$ is the modified Bessel function of order $n$ with argument $\lambda=k_\perp^2w_{\perp s}^2/2\Omega_s^2$, 
\begin{equation}
A_n=\frac{1}{\omega}\frac{T_{\perp s}-T_{\parallel s}}{T_{\parallel s}}+\frac{1}{k_{\parallel} w_{\parallel s}}\frac{(\omega-k_{\parallel}U_s-n\Omega_s)T_{\perp s}+n\Omega_jT_{\parallel s}}{\omega T_{\parallel s}}Z_0,
\end{equation}
\begin{equation}
B_n=\frac{1}{\omega k_{\parallel}}(\omega-k_{\parallel}U_s)+\frac{\omega-n\Omega_j}{k_{\parallel}}A_n,
\end{equation}
\begin{equation}
Z_0(\zeta)=\frac{1}{\sqrt{\pi}}\mathcal P\int\limits_{-\infty}^{\infty}dz\,\frac{e^{-z^2}}{z-\zeta}+i\sqrt{\pi}\,\text{sgn}(k_{\parallel})e^{-\zeta^2},
\end{equation}
and $\zeta=(\omega-k_{\parallel}U_s-n\Omega_j)/k_{\parallel}w_{\parallel s}$.

For the cold plasma case, we use the cold plasma susceptibility accounting for field-aligned relative drifts between the species\cite{Verscharen:2013c}:
\begin{equation}
\underline{\underline{\chi}}_s= \begin{pmatrix}S_s & -iD_s & J_s \\
iD_s & S_s & M_s \\
J_s & -M_s & P_s\end{pmatrix},
\end{equation}
where $S_s=(R_s+L_s)/2$, $D_s=(R_s-L_s)/2$,
\begin{equation}
R_s=-\frac{\omega_{p,s}^2}{\omega^2}\frac{\omega-k_{\parallel}U_s}{\omega-k_{\parallel}U_s+\Omega_s},
\end{equation}
\begin{equation}
L_s=-\frac{\omega_{p,s}^2}{\omega^2}\frac{\omega-k_{\parallel}U_s}{\omega-k_{\parallel}U_s-\Omega_s},
\end{equation}
\begin{equation}
J_s=-\frac{\omega_{p,s}^2}{\omega^2}k_{\perp}U_s\frac{\omega-k_{\parallel}U_s}{\left(\omega-k_{\parallel}U_s\right)^2-\Omega_s^2},
\end{equation}
\begin{equation}
M_s=i\frac{\omega_{p,s}^2}{\omega^2}k_{\perp}U_s\frac{\Omega_s}{\left(\omega-k_{\parallel}U_s\right)^2-\Omega_s^2},
\end{equation}
and
\begin{equation}
P_s=-\frac{\omega_{p,s}^2}{\omega^2}\left[\frac{\omega^2}{\left(\omega-k_{\parallel}U_s\right)^2}+\frac{k_{\perp}^2U_s^2}{\left(\omega-k_{\parallel} U_s\right)^2-\Omega_s^2}\right].
\end{equation}

By using the much more numerically efficient bi-Maxwellian or cold-plasma expressions, rather than direct integration of the momentum derivatives, this `hybrid approach' significantly decreases the computational costs, and is appropriate for cases in which some of the components are accurately treated as cold or a simple Maxwellian, e.g., in evaluating VDFs from hybrid plasma simulations or when some species are only known through their velocity moments rather than their full VDFs.

\section{Eigenfunctions and Heating Rates}
\label{app:eigen}


With the complex frequency $\omega$ determined, the eigenfunctions for the perturbed densities $\delta n_s$, velocities $\delta \V{U}_s$, and electromagnetic fields $\V{E}$ and $\V{B}$ can be calculated through evaluation of the linearized Maxwell's equations, the continuity equation, and the wave equation Eqn.~\ref{eqn:wave_equation}, using the routines implemented in \plume\citep{Klein:2025-PLUME}.
Time scales are normalized to the reference cyclotron frequency $\Omega_\textrm{ref}$ and spatial scales to the reference inertial length $d_{\mathrm{ref}}$.

We choose the complex Fourier coefficient for $\hat{E}_x \equiv  E_x/E_{\perp,1}=1$, where $E_{\perp,1}$ is an arbitrary real constant used to specify the overall amplitude of the linear eigenfunction, and solve for the other components using Eqn.~\eqref{eqn:wave_equation} in terms of 
$E_{\perp,1}$, yielding
\begin{equation}
\hat{E}_{\parallel} \equiv\frac{E_{||}}{E_{\perp,1}}=
\frac{\Lambda_{yx}\Lambda_{zy}-\Lambda_{yy}\Lambda_{zx}}
{\Lambda_{yy}\Lambda_{zz}-\Lambda_{yz}\Lambda_{zy}}
\label{eqn:Epar}
\end{equation}
and
\begin{equation}
\hat{E}_{\perp,2} \equiv \frac{E_{\perp,2}}{E_{\perp,1}}=
-\frac{\Lambda_{zx}+\Lambda_{zz}(E_{||}/ E_{\perp,1})}{\Lambda_{zy}}
=\frac{\Lambda_{zx}\Lambda_{yz}-\Lambda_{zz}\Lambda_{yx}}
{\Lambda_{yy}\Lambda_{zz}-\Lambda_{yz}\Lambda_{zy}},
\label{eqn:Eperp2}
\end{equation}
where $\Lambda_{ij}$ are the elements of the $3 \times 3$ tensor  $\underline{\underline{\Lambda}}$ in Eqn.~\eqref{eqn:wave_equation}.

Combining these solutions for the complex Fourier coefficients of the components of $\V{E}$ with the solutions for the complex frequency $\omega$ and wavevector  $\V{k} = k_\perp \xhat + k_\parallel \zhat$, we  find the complex Fourier coefficients of the  magnetic field eigenfunctions using Faraday's Law, Fourier transformed in time and space, $\omega \V{B}= c \V{k} \times \V{E}$, giving
\begin{equation}
\frac{B_{\perp,1}}{ E_{\perp,1}}= -\frac{ck_\parallel (E_{\perp,2}/E_{\perp,1})}{\omega}=
-\frac{k_\parallel d_{\textrm{ref}}}{(v_A/c) (\omega/\Omega_{\textrm{ref}})}\frac{E_{\perp,2}}{E_{\perp,1}},
\label{eq:Bperp1}
\end{equation}
\begin{equation}
\frac{B_{\perp,2}}{E_{\perp,1}}= \frac{ck_\parallel - ck_\perp (E_{||}/E_{\perp,1})}{\omega}
=-\frac{[k_\perp d_{\textrm{ref}} (E_{||}/E_{\perp,1}) - k_\parallel d_{\textrm{ref}}]}{(v_A/c) (\omega/\Omega_{\textrm{ref}})},
\label{eq:Bperp2}
\end{equation}
and
\begin{equation}
\frac{B_{||}}{E_{\perp,1}}= \frac{ck_\perp (E_{\perp,2}/E_{\perp,1})}{\omega}=
\frac{k_\perp d_{\textrm{ref}}}{(v_A/c) (\omega/\Omega_{\textrm{ref}})}\frac{E_{\perp,2}}{E_{\perp,1}}.
\label{eq:Bpar}
\end{equation}

We use the linearized continuity equation,
\begin{equation}
\frac{\partial \delta n_{s}}{\partial t} + U_s \frac{\partial \delta n_{s}}{\partial z} =
-n_{0s}\nabla \cdot \delta\V{U}_{s},
  \label{eqn:Continuity}
  \end{equation}
   including the normalized equilibrium parallel flow $U_s$, which we can express in terms of the momentum drift for species $s$,
\begin{equation}
    \overline{U}_s \equiv \frac{U_s}{v_{A,\textrm{ref}}} = \frac{P_s}{m_{\textrm{ref}}v_{A,\textrm{ref}}} \frac{m_{\textrm{ref}}}{m_s}
\end{equation}
to solve for the complex Fourier coefficient of the normalized number density fluctuation, $\delta n_{s}/n_{0s}$,
given by 
\begin{multline}
\frac{\delta n_{s}}{n_{0s}} =  \frac{k_\perp \delta U_{xs} + k_\parallel \delta U_{zs}}
{\omega-k_\parallel U_s} \\
=\frac{c}{v_{A,\textrm{ref}}}\frac{k_\perp d_{\textrm{ref}}(\delta \overline{U}_{xs}) + k_\parallel d_{\textrm{ref}} (\delta \overline{U}_{zs})}
{\omega/\Omega_{\textrm{ref}}-k_\parallel d_{\textrm{ref}} \overline{U}_s}\frac{E_{\perp,1}}{B_0},
\label{eqn:density}
\end{multline}
where we normalize the velocity fluctuations as $\delta \overline{\V{U}}_s = \delta \V{U}_s/(c E_{\perp,1}/B_0)$.

We can then determine the perturbed velocity fluctuations for each component $\delta \V{U}_s$ by recognizing that the total current density (including any parallel flow) due is $\V{j}_s = q_s (n_{0s}\,\delta \V{U}_s + \delta n_{s} U_s \zhat)$.
Using the susceptibility tensor to calculate $\V{j}_s$ through Eqn.~\ref{eqn:perturbed_current} yields
\begin{equation}
\delta \V{U}_s = -\frac{i \omega}{4 \pi q_s n_{0s}} \underline{\underline{\chi_s}}(\V{k},\omega) \cdot \V{E} - \frac{\delta n_{s}}{n_{0s}} U_s \zhat.
\label{eqn:Us}
\end{equation}
or
\begin{multline}
    \delta \overline{U}_{x,s}=-\frac{i\omega}{\Omega_{\textrm{ref}}}\left(\frac{v_{A,\textrm{ref}}}{c}\right)^2
    \frac{n_{0,\textrm{ref}}}{n_{0,s}} 
    \frac{q_{\textrm{ref}}}{q_s} 
    \frac{\chi_{xl}E_l}{E_{\perp,1}}
\quad , \quad \\
    \delta \overline{U}_{y,s}=-\frac{i\omega}{\Omega_{\textrm{ref}}}\left(\frac{v_{A,\textrm{ref}}}{c}\right)^2
    \frac{n_{0,\textrm{ref}}}{n_{0,s}} 
    \frac{q_{\textrm{ref}}}{q_s} 
    \frac{\chi_{yl}E_l}{E_{\perp,1}}
\quad , \quad \\
    \delta \overline{U}_{z,s}=
    \frac{
    -\frac{i\omega}{\Omega_{\textrm{ref}}}
    \left(\frac{v_{A,\textrm{ref}}}{c}\right)^2
    \frac{n_{0,\textrm{ref}}}{n_{0,s}} 
    \frac{q_{\textrm{ref}}}{q_s} 
    \frac{\chi_{zl}E_l}{E_{\perp,1}}
    -\frac{{k}_\perp d_{\textrm{ref}} \delta \overline{U}_{x,s}\overline{U}_s}
    {\omega/\Omega_{\textrm{ref}}-k_\parallel d_{\textrm{ref}} \overline{U}_s}}
    {1+ \frac{{k}_\parallel d_{\textrm{ref}}\overline{U}_s }
    {\omega/\Omega_{\textrm{ref}}-k_\parallel d_{\textrm{ref}} \overline{U}_s}}.
\end{multline}

We compute the power emitted or absorbed by each component in the weak damping limit, following the routines implemented in 
\plume\cite{Klein:2025-PLUME}, 
which follow \citet{Stix:1992},
\begin{equation}
    \frac{\gamma_s(\vec{k})}{\omega_{\text{r}}(\vec{k})} = \frac{\vec{E}^*(\vec{k}) \cdot \underline{\underline{\chi}}_s^a(\vec{k}) \cdot \vec{E}(\vec{k})}{4 W_{\text{EM}}(\vec{k})},
\end{equation}
where $\underline{\underline{\chi}}_s^a(\mathbf{k})$ represents the anti-Hermitian component of the susceptibility for species $s$ evaluated at $\gamma=0$, $\vec{E}^*$ represents the complex conjugate of the fluctuating electric field, and 
\begin{equation}
W_{\text{EM}} = \V{B}^*(\V{k})\cdot\V{B}(\V{k}) + \V{E}^*(\V{k})\cdot \frac{\partial}{\partial \omega}[\omega \epsilon_h(\V{k})]\cdot\V{E}(\V{k})
\end{equation}
is the electromagnetic wave energy, where $\epsilon_h$ is the Hermitian part of the dielectric tensor.

\alps\ additionally decomposes the damping governed by the resonance condition  $\omega - k_\parallel v_\parallel -n \Omega_s=0$ into Landau Damping (LD), Transit Time Damping (TTD), and Cyclotron Damping (CD), following the prescription by \citet{Huang:2024}.  
The $n=0$ Landau resonance includes both LD, coupling $E_\parallel$ to a particle's charge,
\begin{multline}
P^{LD}_s =  \frac{i \omega}{16 \pi} \left[ (\chi_{zz,s}^{(n=0)}-\chi_{zz,s}^{(n=0)*})E_{z}E_{z}^* \right. \\
\left. +\chi_{zy,s}^{(n=0)}E_{y}E_{z}^*-\chi_{zy,s}^{(n=0)*}E_{y}^*E_{z})\right]_{\omega=\omega_{\rm r}}, 
\label{eqn:power_ld.hp}
\end{multline}
and TTD, coupling the magnetic field magnitude gradient $\delta B_\parallel$ to the magnetic moment via
\begin{multline}
P^{TTD}_{s} =  \frac{i \omega}{16 \pi} \left[ (\chi_{yy,s}^{(n=0)}-\chi_{yy,s}^{(n=0)*})E_{y}E_{y}^*\right.  \\
 \left.+\chi_{yz,s}^{(n=0)}E_{y}E_{z}^*-\chi_{yz,s}^{(n=0)*}E_{y}^*E_{z})\right]_{\omega=\omega_{\rm r}}. 
\label{eqn:power_ttd.hp}
\end{multline}
In this context, $\chi_{ij}^{n=0}$ is the $ij^{\textrm{th}}$ susceptibility component evaluated for only $n=0$. 

CD occurs via $n\neq 0$ resonances due to forces by $E_\perp$ on the charged particles. 
The power absorption due to CD of the $n^{\rm th}$ harmonic is
\begin{multline}
P_s^{\rm CD,n}=  \frac{\omega}{8 \pi}
\left[|E_{x}|^2\left(\chi_{xx,s}^{n} -\chi_{xx,s}^{n,*}\right) 
+ |E_{y}|^2\left(\chi_{yy,s}^{n} -\chi_{yy,s}^{n,*}\right) \right. \\
\left.+  \left(E_{x}^*E_{y} - E_{y}^*E_{x}\right)\left(\chi_{xy,s}^{n} -\chi_{yx,s}^{n,*}\right)
\right]_{\omega=\omega_{\rm r}}. 
\label{eqn:power_cd.hp}
\end{multline}
As currently implemented, \alps\ only outputs the CD power absorption/emission associated with the $n=\pm 1$ resonance.

\section{The Limits of Analytic Continuation}
\label{app:cheb}

\begin{figure*}
    \centering
    \includegraphics[width=0.3\linewidth]{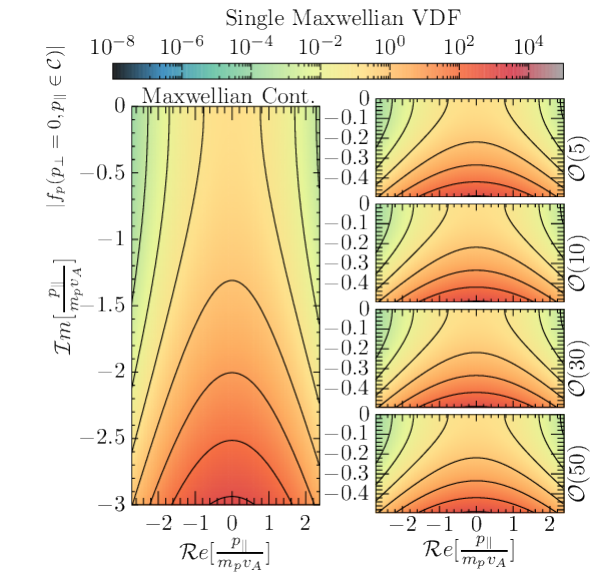}
    \includegraphics[width=0.3\linewidth]{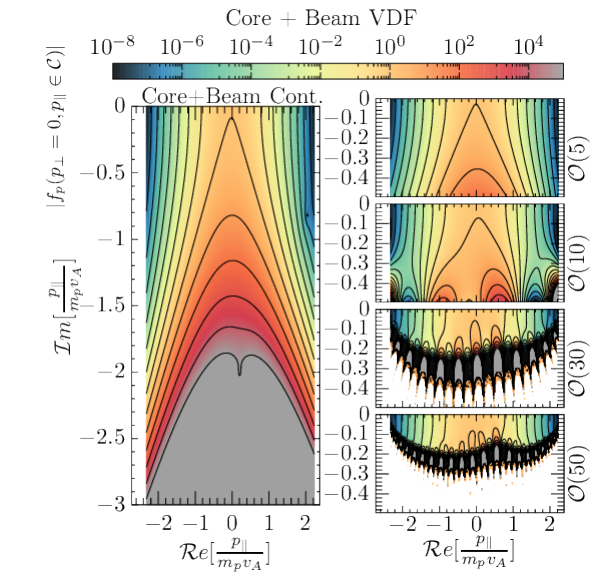}
    \includegraphics[width=0.3\linewidth]{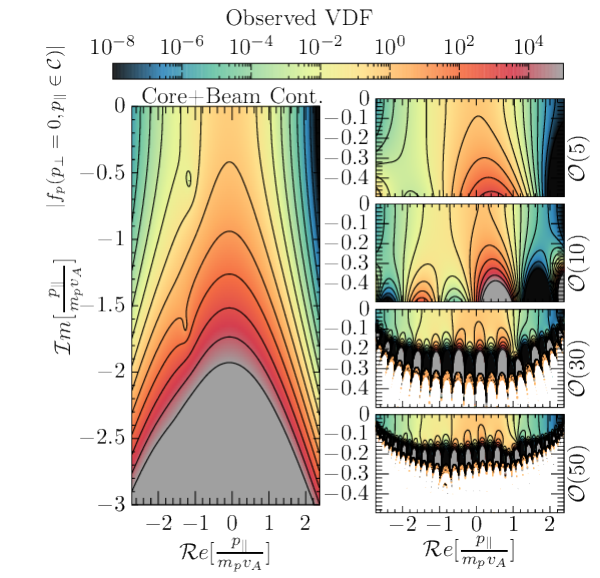}
    \caption{The amplitude of $|f_p|$ evaluated at complex parallel momentum for $p_\perp=0$ using one or two bi-Maxwellians, as well as for several orders of GLLS (5 through 50; right).
    The input VDFs are:
    (left) a single Maxwellian,
    (center) the core-and-beam bi-Maxwellian from \S~\ref{ssec:2-vdfs}, and 
    (right) the observed distribution from \S~\ref{ssec:2-vdfs}.}
    \label{fig:continuation_bM}
\end{figure*}
\begin{figure}
    \centering
    \includegraphics[width=0.9\columnwidth]{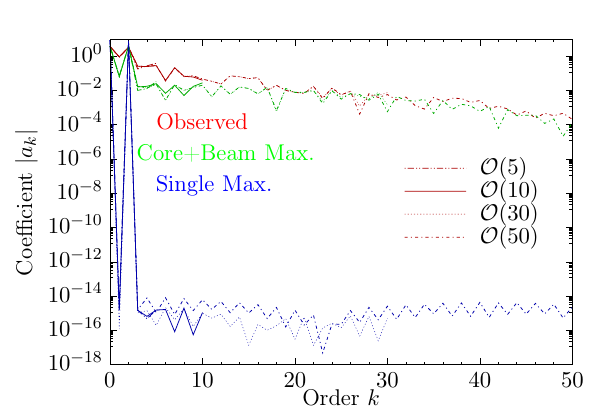}
    \caption{Magnitude of coefficients of GLLS fits $a_k$ for the three VDFs illustrated in Fig.~\ref{fig:continuation_bM}.
    }
    \label{fig:coeffs}
\end{figure}

Given the novel GLLS representation for the analytic continuation, it is important to quantify how far into imaginary parallel momentum space the representation can be accurately and reliably extended.
Towards this end, we consider the continuations of three VDFs: a single Maxwellian, the best fit core-and-beam bi-Maxwellian $f_{C\&B}(p_\perp=0)$ used in \S~\ref{ssec:2-vdfs},  and the observed VDF $f_{Obs.}(p_\perp=0)$ also described in \S~\ref{ssec:2-vdfs}.
We evaluate the amplitude of the complex valued representation of the VDF $|f_p[\mathbb{R}e(p_\parallel),\mathbb{I}m(p_\parallel);p_\perp=0]$ for all three models. 
For the GLLS representation, we apply  orders 5, 10, 30, and 50.
These comparisons are shown in Fig.~\ref{fig:continuation_bM}, and the coefficients for the GLLS fits are shown in Fig.~\ref{fig:coeffs}.

All orders of the GLLS method yield the quantitatively identical continuation behavior as the single Maxwellian representation when the underlying VDF is a single Maxwellian.  
This can be understood by noting that only two of the coefficients, $a_0$ and $a_2$, are effectively non-zero for all orders, blue lines in Fig.~\ref{fig:coeffs}, regardless of the order of the GLLS representation.
This matches with our intuitive expectations, as the log of a Maxwellian is a second-order polynomial.
This simple quadratic expression matches the single Maxwellian well, and the lack of additional terms in the representation produces the same complex structure at larger values of $|\mathbb{I}m(p_\parallel)|$.

For the core-and-beam VDF, a broader range of orders have non-negligible power.
These additional terms lead to oscillations in the complex GLLS representation of $f_p$, and thus increased amplitudes of $|f_p|$, at larger $|\mathbb{I}m(p_\parallel)|$ values compared to the relatively shallower increase in $|f_p|$ when evaluating the sum of two Maxwellians in $f_{C\&B}$.
This trend is further exacerbated for the observed distribution $f_{Obs.}$, where the non-Maxwellian structure in the VDF leads to more power in the lower-order coefficients $a_k$ compared to the either of the simpler VDFs.

This increase in $|f_p|$  impacts the resolution of strongly damped waves, as the associated values of $|\mathbb{I}m(p_\parallel)|$ for such modes are large enough for the numerical evaluation of the analytic continuation of $f_p$ to be inaccurate.
\change{As seen in the right-hand panels of Fig.~\ref{fig:continuation_bM}, there is a larger increase in $|f_p|$ with higher order representations for more strongly damped solutions.
For this reason, users should select the lowest-order representation sufficient to capture the VDF structure. This approach not only optimizes the numerical efficiency of this scheme, but it also avoids using unnecessarily high orders which may introduce numerical artifacts for strongly damped solutions.
Overall, given} the superior behavior of the GLLS method for weakly damped solutions, we find using the Chebyshev polynomials rather than a low-order Maxwellian fit is the optimal choice for non-Maxwellian VDFs. 

\bibliography{master}

\end{document}